# Noise-squeezed forward Brillouin lasers in multimode fiber microresonators

Mingming Nie[1,*], Kunpeng Jia[2], Shining Zhu[2], Zhenda Xie[2] and Shu-Wei Huang[1,*]

[1]*Department of Electrical, Computer and Energy Engineering, University of Colorado Boulder, Boulder, Colorado 80309, USA*
[2]*National Laboratory of Solid State Microstructures, School of Electronic Science and Engineering, College of Engineering and Applied Sciences, School of Physics, and Collaborative Innovation Center of Advanced Microstructures, Nanjing University, Nanjing 210093, China*
[*]*Corresponding author: mingming.nie@colorado.edu, shuwei.huang@colorado.edu*

**Abstract:** Stimulated Brillouin scattering (SBS) in low-power and compact microresonators has created a new field in cavity nonlinear photonics due to the marriage between acoustic and optical signal processing. Considering the fundamental differences between backward SBS and forward SBS processes, it is challenging to observe the coexistence of both processes in the same microresonator, as well as the photon noise suppression for the forward stimulated Brillouin laser (FSBL). In this paper, we demonstrate the first 20-dB-noise-squeezed FSBL generation excited by the coexisting backward SBL (BSBL) in an ultrahigh-quality-factor Fabry–Pérot (FP) microresonator based on multimode fiber (MMF). Multiple FSBLs and BSBLs are cascaded by multiple intermodal SBS processes in the multimode microresonator, where the cascaded process between backward SBS and forward SBS process (pump-BSBL-FSBL) provides a route towards additional noise squeezing, rendering the FSBL phase noise to be -120 dBc/Hz at 1 MHz offset frequency. Furthermore, we demonstrate the first Brillouin-Kerr soliton from a high-order BSBL, which also coexists with FSBLs. Our experimental results show the potential of MMF FP microresonator as an ideal testbed for high-dimensional nonlinear cavity dynamics and laser source with ultrahigh coherence.

## Introduction

Stimulated Brillouin scattering (SBS) is a fundamental interaction between light and travelling acoustic waves, which arises primarily from electrostriction and photoelastic effects. Due to the low size, weight, power and cost, SBS in microresonators has arguably created a new field in cavity nonlinear photonics thanks to the combination between photonics provided broad bandwidth and acoustics enabled fine spectral resolution [1].

According to the propagation direction relative to the pump, forward SBS (FSBS) and backward SBS (BSBS) can be identified. The backward stimulated Brillouin laser (BSBL) in microresonators has emerged as a promising coherent laser source, achieving sub-hertz fundamental linewidth [2–5]. In particular, by utilizing the interplay between BSBL and cavity Kerr nonlinearity, the recently demonstrated Brillouin-Kerr soliton microcomb [3,6,7] enables the photonic flywheel with ultranarrow comb linewidth and ultralow soliton jitter, which are key to the success of many intriguing applications, such as optical atomic clocks [8–10], optical gyroscopes [11–14] and microwave photonics [2,15–17].

On the contrary, FSBS process is fundamentally different from BSBS process, in terms of phase matching, acoustic wave properties and microresonator requirement [1]. The phase matching of FSBS process is fulfilled between two co-propagating optical waves and near-zero-wavenumber acoustic wave. The transverse guiding of the acoustic wave necessitates delicately tailored microresonator geometry, while the low-frequency phonon requires a multimode microresonator for intermodal FSBS process due to the large free spectral range (FSR). Moreover, the guided acoustic wave usually leads to large phonon lifetime and phonon noise squeezing [18–20], triggering non-reciprocal applications. Considering these fundamental differences, it is challenging to observe the coexistence of FSBS and BSBS processes in the same microresonator, as well as the photon noise suppression for the forward stimulated Brillouin laser (FSBL), despite the potential of interesting phenomena, improved performances, and new applications.

Fiber is a good candidate to study both the BSBS and FSBS process due to the relatively large Brillouin gain coefficient and the cylindrical geometry supporting transversely oscillating acoustic waves [21]. BSBL with narrow linewidth is first demonstrated in a single-mode fiber cavity since 1980's [22] and recently a fiber Fabry–Pérot (FP) microresonator with ultrahigh quality (Q) factor [3,6]. On the other hand, FSBS has been well studied in various fibers [21,23–28] and applied in applications such as sensing [29–31] and mode-locked lasers [32,33]. Moreover, FSBL is recently demonstrated in a panda-type polarization maintaining fiber laser [26]. Despite the theoretical prediction, noise-squeezed FSBL has not been achieved.

In this paper, we demonstrate the first 20-dB-noise-squeezed FSBL generation excited by the coexisting BSBL in an ultrahigh-Q multimode fiber (MMF) FP microresonator with increased photon lifetime and reduced phonon lifetime. Multiple FSBLs and BSBLs are cascaded by multiple intermodal SBS processes in the multimode microresonator, where the cascaded process between BSBS and FSBS process (pump-BSBL-FSBL) provides a route towards additional noise squeezing, rendering the FSBL phase noise to be -120 dBc/Hz at 1 MHz offset frequency. Furthermore, we demonstrate the first Brillouin-Kerr soliton from a high-order BSBL, which also coexists with FSBLs. Our experimental results show the potential of MMF FP microresonator as an ideal testbed for high-dimensional nonlinear cavity dynamics and laser source with ultrahigh coherence, which are critical for acoustic and optical signal generation and processing.

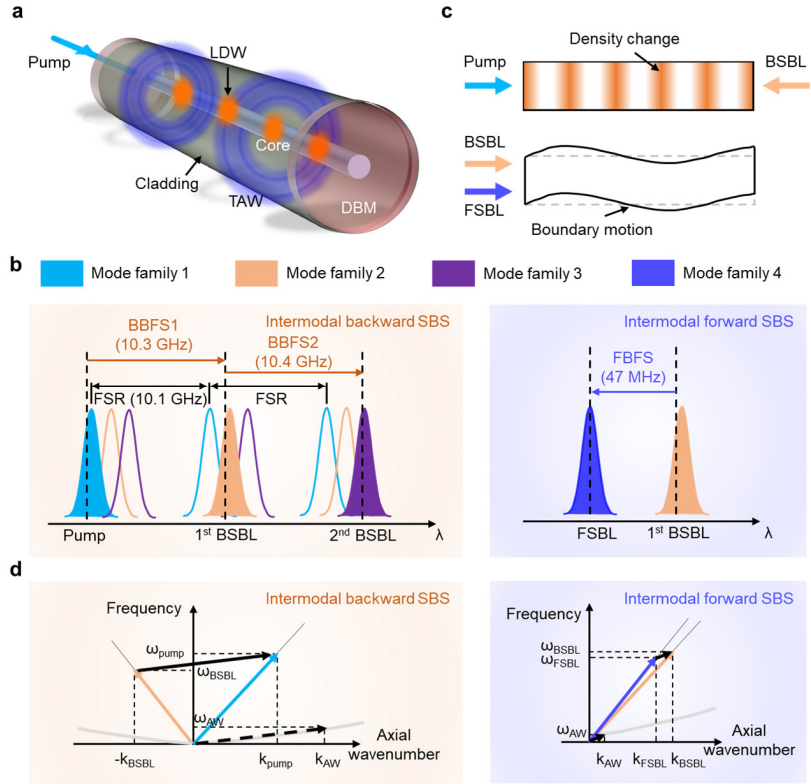

**Fig. 1. Principle of operation.** (a) Schematic diagram of BSBL and FSBL generation in a MMF FP microresonator LDW: longitudinal density wave, TAW: transverse acoustic wave, DBM: dielectric Bragg mirror. (b) Illustration of resonant pump, 1st BSBL and 2nd BSBL via intermodal BSBS process (left) and resonant FSBL via intermodal FSBS process pumped by 1st BSBL (right). (c) Optomechanical effect in the BSBS (top) and FSBS (bottom) process. (d) Phase matching diagram for intermodal BSBS (left) and FSBS (right) process.

## Results

**Principle of operation.** Figure 1a shows the schematic diagram of BSBL and FSBL generation in a MMF FP microresonator (see Methods for details). To intermodally excite BSBLs and FSBLs in the microresonator, three conditions must be satisfied: (i) difference between Brillouin frequency shift (BFS) and the microresonator FSR; (ii) good overlapping between the Brillouin mode resonance and Brillouin gain spectrum; (iii) high Qs for both modes to lower the lasing threshold.

As shown in Fig. 1b, the microresonator FSR is 10.079-GHz, which is slightly smaller than the experimental backward Brillouin frequency shifts (BBFSs) ranging from 10.3 GHz to 10.6 GHz and much larger than the forward Brillouin frequency shift (FBFS) at ~100 MHz level. The loaded Qs of five typical spatial modes including the pump mode (MMF fundamental mode) are measured to be ultrahigh approaching $4\times10^8$ [3]. The Qs for all the supported (~100) modes in the MMF are believed to be larger than $1\times10^8$ [3]. Multiple BSBLs and FSBLs are feasible due to the flexible phase matching between a great variety of high-Q optical modes and acoustic modes in the MMF.

The BSBL is excited through the longitudinal density wave in the core, while the FSBL is excited through the boundary motion of the core induced by the oscillating transverse acoustic wave (Fig. 1c). Of note, here the FSBL is excited by the intra-microresonator BSBL. The phase matching diagrams for BSBL and FSBL generation are shown in Fig. 1d, where the wavenumbers of the acoustic waves are $\sim10^7$ and $\sim0$ respectively [1].

**FSBS gain spectrum measurement for MMF.** Before the study of FSBL generation in the microresonator, the FSBS gain spectrum of our MMF is measured via pump-probe technique [34,35]. The pump laser is centered at 1560 nm with a 2-ns pulse width and a repetition rate of 87 kHz

(period of 11.5 μs, see Supplementary Information for details), while the probe laser is a continuous wave (CW) single-frequency laser centered at 1543 nm. A Mach-Zehnder interferometer (MZI) is introduced to detect the probe phase change induced by the FSBS process. In one arm, the probe is unmodulated, while in the other arm the probe is phase modulated by the co-propagating high-peak-power pump through the stimulated oscillations of radially guided acoustic modes. The pump and probe laser are combined via a dichromatic mirror and injected into the 30-m-long MMF. At the output of the MMF, the pump and probe are separated by another dichromatic mirror. Other experimental details can be found in Methods.

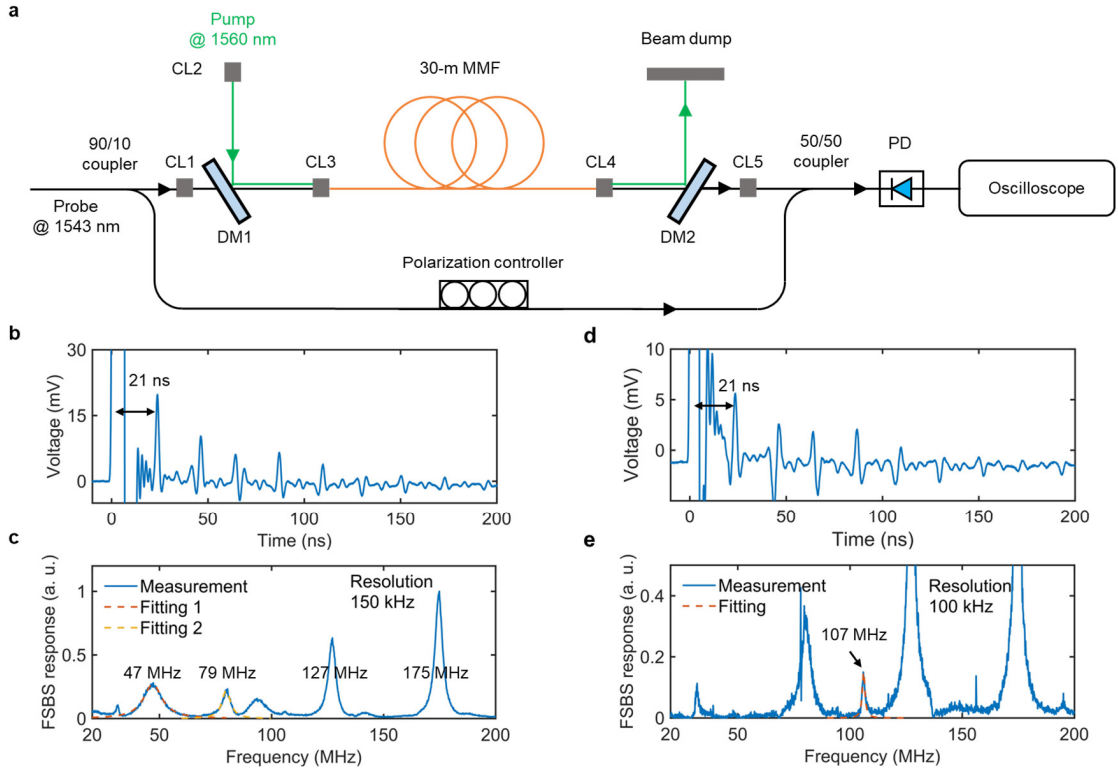

**Fig. 2. FSBS gain spectrum measurement for MMF.** (a) Experimental setup for MMF FSBS gain spectrum measurement. CL: collimating lens, DM: dichromatic mirror, PD: photodetector. (b)(d) Temporal response of the probe laser. (c)(e) Normalized FSBS gain spectrum.

In time domain (Fig. 2b), the probe laser first undergoes cross phase modulation (XPM) from the pump pulse via the instantaneous Kerr effect and later phase modulation through FSBS process dominates and lasts for hundreds of nanoseconds. The impulses result from the successive reflections of the guided acoustic wave at the outer boundary of the cladding, whose temporal separation of ~21 ns is determined by the cladding diameter of 125 μm and the acoustic velocity of 6,010 ± 50 m/s in silica [36]. Figure 2c shows the normalized FSBS gain spectrum of our MMF by converting the temporal response of the phase modulation into the power spectrum in Fourier domain. Due to the multiple mode excitation for both the pump and probe laser, multiple FSBS process including intramodal and intermodal FSBS can occur, leading to a serial of resonance peaks at 47 MHz, 79 MHz, 127 MHz and 175 MHz. By changing the coupling angles of the pump and probe laser therefore different excited spatial modes, another resonance peak at 107 MHz can also be observed (Figs. 2d and 2e). Here we focus on the peaks at 47 MHz, 79 MHz and 107 MHz with corresponding fitted linewidths of 10 ± 0.3 MHz, 4.9 ± 0.3 MHz and 1.4 MHz ± 0.4 MHz, since they are related to the experimental results in the next sections. Of note, the relatively large acoustic resonance linewidths are mainly caused by the acoustic impedance resulting from the glue outside the cladding [29] (see Methods), while the linewidth differences are attributed to the different boundary motions induced by mode-dependent radial pressure in the MMF.

**FSBL excited by BSBL.** Figure 3a shows the experimental setup for generating and characterizing the BSBL and FSBL in the MMF FP microresonator. When 0.8-W pump laser (ECDL, Toptica CTL 1550) is loaded into the microresonator, BSBLs with multiple cascaded orders can be generated via the intermodal BSBS process (Figs. 3b, 3c and 3d). Cascaded orders with different BFSs ranging from 10.3

GHz to 10.6 GHz are achieved by tuning the pump center wavelength.

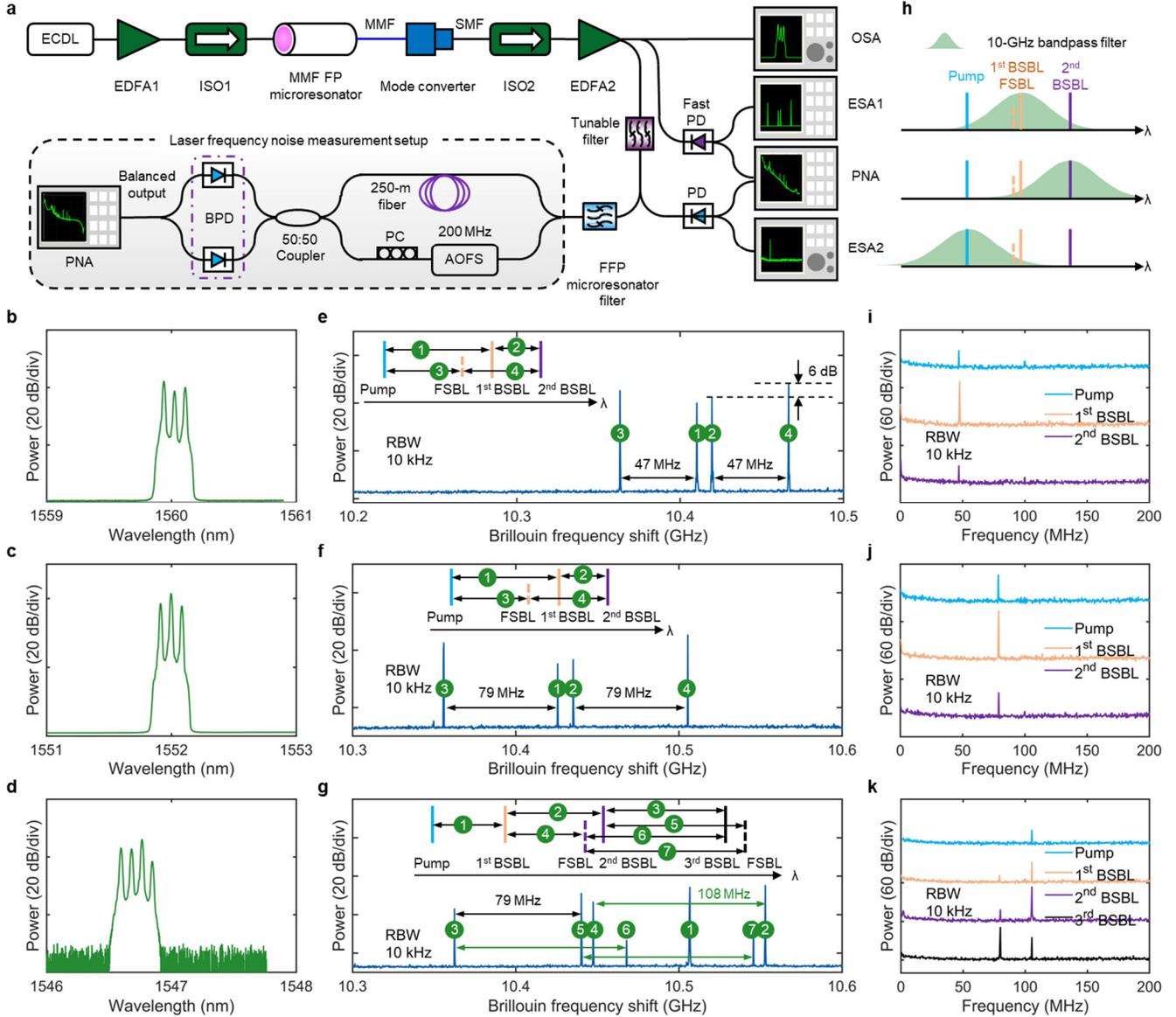

**Fig. 3. Multiple BSBLs and FSBLs generation in the MMF FP microresonator.** (a) Experimental setup for BSBL and FSBL generation in the MMF FP microresonator and corresponding characterization measurement. ECDL: external-cavity diode laser, EDFA: Erbium-doped fiber amplifier, ISO: optical isolator, SMF: single-mode fiber, OSA: optical spectrum analyzer, PD: photodetector, ESA: electrical spectrum analyzer, PNA: phase noise analyzer, AOFS: acousto-optic frequency shifter, PC: polarization controller, BPD: balanced photodetector. (b-d) Optical spectrum showing cascaded BSBL generation. (e-g) RF spectrum showing the beat note between different optical frequencies, labelled by numbers. Main peaks showing the BBFSs are labelled by 1 and 2 in (e), 1 and 2 in f and 1, 2 and 3 in (g). Other additional peaks showing the beat notes between FSBLs and either the pump or BSBLs, are labelled by the sequential numbers. Inset: frequency relationship between the pump, cascaded BSBLs and FSBLs. (h) Schematic of optical frequency filtering via a narrowband filter. (i-k) RF spectrum showing the modulation frequencies from filtered different cascaded BSBLs. The direct components (DCs) from the PD are kept to be the same for the filtered optical frequencies in the measurement. Therefore, the filtered optical frequency corresponding to the highest RF peak is the main reason for the modulation. Case A: (b)(e)(i), Case B: (c)(f)(j), Case C: (d)(g)(k).

Interestingly, besides the main BFS peaks representing the BBFSs, additional RF peaks with equal distance of 47 MHz (Fig. 3e), 79 MHz (Fig. 3f) and 108 MHz (Fig. 3g) appear simultaneously in a specific pump frequency tuning range, indicating optical modulation and new optical frequency generation inside the microresonator. In order to examine the modulation origin, pump and cascaded BSBLs are separately filtered out via a tunable filter with 3-dB bandwidth of 10-GHz (Fig. 3h). As shown in Figs. 3i, 3j and 3k, the optical modulations at 47 MHz (Fig. 3e), 79 MHz (Fig. 3f), 79 MHz (Fig. 3g) and 108 MHz (Fig. 3g) originate from 1st BSBL, 1st BSBL, 3rd BSBL and 2nd BSBL, respectively.

Considering the measured FSBS gain spectrum in Figs. 2c and 2e, we believe the optical modulation is caused by the FSBS induced phase modulation. In the insets of Fig. 3e, 3f and 3g, the frequency relationship between the pump, the cascaded BSBLs and FSBLs are plotted. The optical frequency beat notes exactly match the measured RF peaks, confirming that FSBLs are excited by BSBLs. Only single-sideband (SSB) modulations are observed for all the three cases due to the critical overlapping between the FSBS gain spectrum and narrowband mode resonance. Both the laser-frequency-dependent and SSB modulation behavior exclude the conventional mechanical breathing mode as the dominant reason for the observed phenomena.

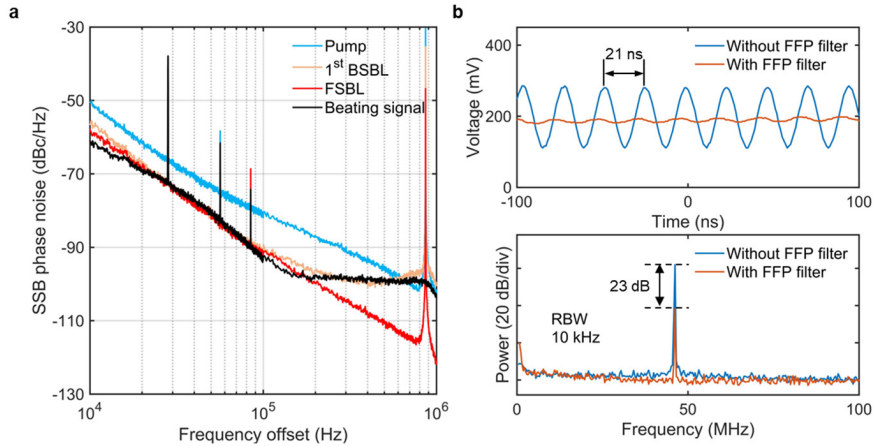

**Fig. 4. Phase noise measurement.** (a) SSB phase noise spectra of pump, 1st BSBL, FSBL and beat note between 1st BSBL and FSBL. The coherent artefact at 0.85 MHz result from the 250-m delay fiber used in the frequency discriminator []. (b) Temporal traces (top) and RF spectrum (bottom) from the beat note between 1st BSBL and FSBL, without (blue) and with (orange) FFP filter.

**Noise-squeezed FSBL generation.** In ultrahigh-Q microresonator it usually exhibits laser phase noise suppression for the generated BSBL since the intracavity photon decaying rate of ~MHz level is much slower than the phonon decaying rate of ~ 100 MHz so that the photon noise can be quickly damped to the phonon through the BSBS process. Similarly, for Case A in Fig. 3b the generated FSBL is expected to be noise-squeezed since the photon decaying rate of ~1 MHz is smaller than the intermodal FSBS decaying rate of ~10 MHz (Fig. 2c).

As shown in Fig. 4a, the SSB phase noises spectra of the pump, 1st BSBL and FSBL are measured via delayed self-heterodyned method [37]. A 10-GHz tunable filter is introduced to filtered out the pump and the combination of 1st BSBL and FSBL, respectively. The SSB phase noise of the pump is shown in Fig. 4a with a blue line. As the power of FSBL is 6 dB larger than the 1st BSBL (Fig. 3e), the measured phase noise from the 10-GHz filter output is believed to belong to the FSBL (red line in Fig. 4a). Due to the small FBFS of 47 MHz, an additional optical filter made of a fiber FP (FFP) microresonator with 3-dB linewidth of 0.7 MHz (see Supplementary Information) is employed to reduce the contrast between the FSBL and 1st BSBL by 23 dB (Fig. 4b) and filter out the 1st BSBL for phase noise measurement (see Methods for details). Compared with the pump, the phase noise of 1st BSBL (orange line in Fig. 4a) is ~10 dB lower, verifying the linewidth narrowing effect in the BSBS process. In addition, the FSBL shows phase noise reduction starting from 100 kHz offset frequency compared to its pump (1st BSBL). At offset frequency of 1 MHz, the noise of FSBL is -120 dBc/Hz, representing 20 dB suppression, which is determined by the square of the ratio ($10^2$=100) between the photon and phonon decaying rate [38]. The phase noise of beat note between FSBL and 1st BSBL at 47 MHz (black line in Fig. 4a) also indicates FSBL has smaller phase noise than 1st BSBL (see Supplementary Information). To the best of our knowledge, this is the first experimental result demonstrating cascaded process between BSBS and FSBS (pump-BSBL-FSBL) and noise squeezing in FSBL generation, which can benefit critical applications requiring ultrahigh coherence and dual frequencies with small frequency shift such as dual-frequency laser interferometer [39,40]. Of note, although FSBL is excited by the 1st BSBL (Fig. 3b), its appearance has little effect on the other intracavity process. In Supplementary Information, the phase noises of 1st BSBL before and after FSBL generation are found to be the same, indicating independent FSBS process.

**Coexistence of FSBL and Brillouin-Kerr soliton.** Figure 5 shows the coexistence of dissipative Kerr soliton (DKS) and FSBLs generated by high-order BSBLs. Both the spectrum of the 2-soliton perfect crystal with sech$^2$ profile (Fig. 5a) and the clean beat note of the comb repetition rate (inset of Fig. 5a) confirm the stable and low-noise mode-locking state. The zoomed-in spectrum (Fig. 5b) suggests the soliton pump originates from 3rd BSBL, which is the first Brillouin-Kerr soliton excited by a high-order BSBL. Other comb states including the single soliton, soliton crystals and soliton coexisting with pump Turing patterns can be found in Supplementary Information.

Following the steps in the previous sections, two FSBLs are examined to be generated by 2nd BSBL and 3rd BSBL with FBFSs of 108 MHz and 47 MHz, respectively (Fig. 5c). Despite the same origin from 3rd BSBL, the DKS generation is independent of FSBL generation. No modulation peak is observed in the filtered DKS microcomb (Fig. 5d) due to the independent nonlinear processes.

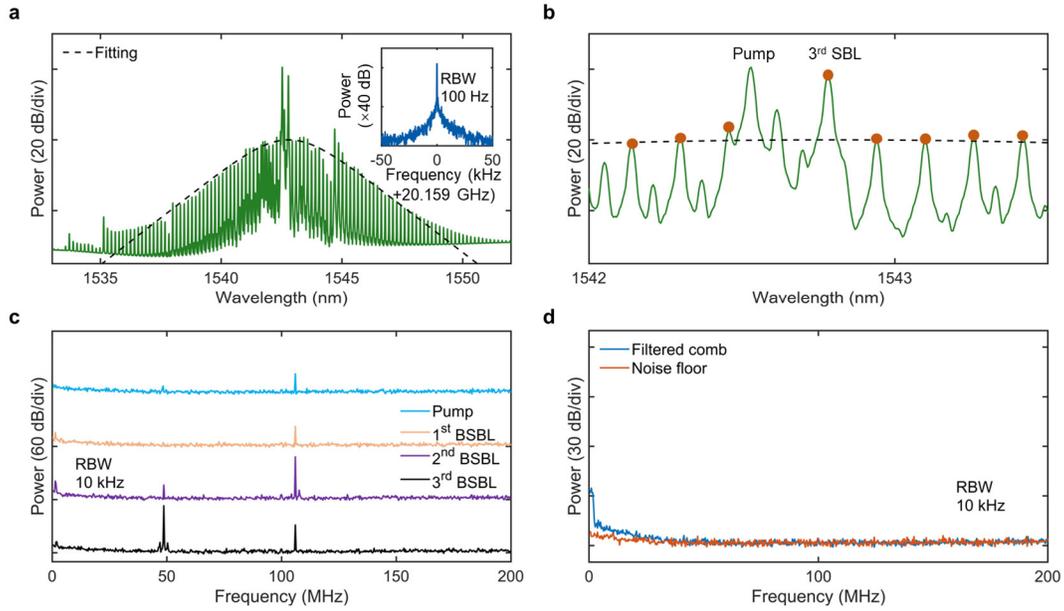

**Fig. 5. Coexistence of FSBL and BSBL-DKS.** (a) Optical spectrum of the 2-soliton perfect soliton crystal. Inset: RF beat note of the comb repetition rate. (b) Zoomed-in spectrum. The orange dots show the comb line positions. (c) RF spectrum showing the modulation frequencies from filtered different cascaded BSBLs. (d) RF spectrum of the filtered comb.

## Discussion

In summary, we demonstrate the first 20-dB-noise-squeezed FSBL generation excited by the coexisting BSBL in an ultrahigh-Q MMF FP microresonator. Multiple FSBLs and BSBLs are cascaded by multiple intermodal SBS processes in the multimode microresonator, where the cascaded process between BSBS and FSBS process (pump-BSBL-FSBL) provides a route towards additional noise squeezing, rendering the FSBL phase noise to be -120 dBc/Hz at 1 MHz offset frequency. Furthermore, we demonstrate the first Brillouin-Kerr soliton from a high-order BSBL, which also coexists with FSBLs. Our experimental results show the potential of MMF FP microresonator as an ideal testbed for high-dimensional nonlinear cavity dynamics and laser source with ultrahigh coherence, which are critical for acoustic and optical signal generation and processing.

## Methods

**MMF FP microresonator.** Our ultrahigh-Q MMF FP mesoresonator is fabricated through three steps: (i) Commercial MMF (GIF50E, Thorlabs) with removed coatings is carefully cleaved and encapsuled in a ceramic fiber ferrule. Epoxy is used to firmly glue the fiber in the ferrule. (ii) Both fiber ends are mechanically polished to sub-wavelength smoothness; (iii) both fiber ends are coated with optical DBR with reflectivity over 99.9% from 1530 to 1570 nm.

**FSBS gain spectrum measurement.** The 30-m MMF is coating removed and glued to a fiber holder in order to mimic the state inside the ferrule of the microresonator. The input power for the pump and probe laser are 200 mW and 20 mW, respectively. The small phase modulation (much smaller than $2\pi$) is converted to intensity modulation through the MZI. The polarization controller in the unmodulated arm is used to maximize the output power. The temporal response of FSBS process is sampled by a digitizing oscilloscope and is averaged by 2000 times. Before converted to frequency domain, the temporal trace is gated to remove the XPM induced strong pulse [35]. The normalized FSBS spectrum is then determined by the ratio between the RF power spectrum of the gated trace and the RF power spectrum of the pump pulse.

**1st BSBL filtering using a FFP filter.** Since FSBL excited by 1st BSBL is blue shifted from 1st BSBL by 47 MHz (Fig. 3e), we use a FFP filter with 0.7 MHz linewidth to filter out the 1st BSBL. First, the total power of FSBL and 1st BSBL are amplified to be 100 mW. Second, the temperature of the FFP filter is fine tuned to load the 1st BSBL into the FFP filter. Since the 1st BSBL can be thermally locked at the bule side of the FFP, the output of the FFP filter remains stable, ready for measurement. Of note, the laser amplification and the FFP filter will not change the laser phase noise.

## Data availability
All data generated or analyzed during this study are available within the paper and its Supplementary Information. Further source data will be made available on reasonable request.

## Code availability
The analysis codes will be available on reasonable request.

## Acknowledgments
We thank Professor Scott A. Diddams from CU Boulder and Peter T. Rakich from Yale University for fruitful discussions. M.N. and S.W.H. acknowledge the support from the University of Colorado Boulder and National Science Foundation (OMA 2016244 and ECCS 2048202). K.J., S. Z. and Z.X. acknowledge the support by the National Key R&D Program of China (2019YFA0705000, 2017YFA0303700), Key R&D Program of Guangdong Province (2018B030329001), Guangdong Major Project of Basic and Applied Basic Research (2020B0301030009), Zhangjiang Laboratory (ZJSP21A001), Leading-edge technology Program of Jiangsu Natural Science Foundation (BK20192001) and National Natural Science Foundation of China (51890861, 11690031, 11621091, 11627810, 11674169, 91950206).

## Author contributions
M.N. and S.W.H. conceived the idea of the experiment. M.N. designed and performed the experiment. K.J., S.Z. and Z.X. designed and fabricated the microresonator. M.N. and S.W.H. conducted the data analysis and wrote the manuscript. S.W.H. led and supervised the project. All authors contributed to the discussion and revision of the manuscript.

## Competing interests
The authors declare no competing interest.